\documentclass[12pt,groupedaddress, showpacs, aip, jmp]{revtex4-1}

\usepackage{hyphenat}
\usepackage{bm}
\usepackage{graphicx}
\usepackage{epsfig}
\usepackage{amsmath}
\usepackage{amssymb}
\usepackage{amsthm}
\usepackage{paralist}
\usepackage{subfig}

\DeclareGraphicsExtensions{.eps}

\begin{document}

\title{Wave functions of log-periodic oscillators}
\author{V. Bessa}
\email[Electronic mail(Corresponding author): ]{vagner@fisica.ufc.br}
\author{I. Guedes}
\email{guedes@fisica.ufc.br}
\affiliation{Universidade Federal do Cear\'{a}}

\begin{abstract}

We use the Lewis and Riesenfeld invariant method [\textit{J. Math. Phys.} \textbf{10}, 1458
(1969)] and a unitary transformation to obtain the exact Schr\"{o}dinger wave functions for
time-dependent harmonic oscillators exhibiting log-periodic-type behavior. For each
oscillator we calculate the quantum fluctuations in the coordinate and momentum as
well as the quantum correlations between the coordinate and momentum. We observe
that the oscillator with $m=m_0t/t_0$ and $\omega= \omega_0t_0/t$, which exhibits an exact log-periodic
oscillation, behaves as the harmonic oscillator with $m$ and $\omega$ constant.

\end{abstract}

\pacs{03.65.-w, 03.65.Aa, 03.65.Ge, 03.65.Ca}

\maketitle

\section{\label{sec1}Introduction}
Several types of time-dependent oscillators have been studied along the past years. Examples are: (i) the harmonic oscillator\cite{1}; (ii) the pseudo-harmonic oscillator\cite{1, 2}; (iii) the parametric oscillator\cite{3}; and (iv) the inverted harmonic oscillator\cite{4}. Recently, another interesting class of time-dependent oscillators, named log-periodic oscillators, was studied\cite{5}. 

In Ref. \onlinecite{5}, \"{O}zeren\cite{5} considered the time evolution of five different one-dimensional classical oscillators. The coherent states for each system were constructed by using the SU(1, 1) algebra and their time evolution was investigated.

In this work, we use the Lewis and Riesenfeld\cite{6} (LR) invariant method and a unitary transformation to obtain the exact Schr\"{o}dinger wave function for three out of the five log-periodic-type oscillators investigated by \"{O}zeren\cite{5}, namely: (i) $m(t)=m_0\frac{t}{t_0}$ and $k(t)=k_0\frac{t_0}{t}$; (ii) $m(t)=m_0$ and $k(t)=k_0\left(\frac{t_0}{t}\right)^{2}$; (iii) $m(t)=m_0\left(\frac{t}{t_0}\right)^{2}$ and $k(t)=k_0$. In all three cases $\omega(t)=\omega_0\frac{t_0}{t}$.

The wave functions $\psi_n (q,t)$ for the time dependent harmonic oscillator ($H(t)=\frac{p^2}{2m(t)}+\frac{1}{2}m(t)\omega^2(t)q^2$) obtained in Ref. \onlinecite{1} are written in terms of $\rho$, a c-number quantity satisfying the generalized Milne-Pinney equation ($\ddot{\rho}+\gamma(t)\dot{\rho}+\omega^2(t)\rho=\frac{1}{m^2(t)\rho^3}$), whose solution can be found following the procedure reported in Refs. \onlinecite{7, 8}.
Here we write the solution of the Milne-Pinney equation for each system to obtain the exact wave functions for the oscillators. This paper is outlined as follows. In Sec. \ref{sec2} we briefly review the LR invariant method for the time-dependent harmonic oscillator. In Sec. \ref{sec3} we obtain the wave functions for the oscillators considered, and calculate the correlations between position and momentum and the uncertainty product. For oscillator (i) we construct the coherent states, while for oscillators (ii) and (iii) we construct the squeezed states. The analysis of the phase diagram for the three oscillators is also presented.  Finally, some concluding remarks are added in Sec. \ref{sec4}.

\section{\label{sec2}THE LEWIS AND RIESENFELD INVARIANT METHOD - WAVE FUNCTIONS FOR A TIME-DEPENDENT HARMONIC OSCILLATOR}
Consider a time-dependent harmonic oscillator described by the Hamiltonian

\begin{equation}\label{1}H(t)=\frac{p^2}{2m(t)}+\frac{1}{2}m(t)\omega^2(t)q^2,\end{equation}whose mass ($m(t)$) and angular frequency ($\omega(t)$) depend on time explicitly, and the variables $q$ and $p$ are canonical coordinates with $[q,p]=i\hbar$. From Eq. (\ref{1}), we obtain the equation of motion

\begin{equation}\label{2}\ddot{q}+\gamma(t)\dot{q}+\omega^2(t)q=0,\end{equation}where

\begin{equation}\label{3}\gamma(t)=\frac{d}{dt}\ln{m(t)}.\end{equation}

It is well known that an invariant for Eq. (\ref{1}) is given by\cite{6}

\begin{equation}\label{4}I=\frac{1}{2}\left[ \left(\frac{q}{\rho}\right)^2+(\rho p-m\dot{\rho}q)^{2} \right] \end{equation}where $q(t)$ satisfies Eq. (\ref{2}) and $\rho(t)$ satisfies the generalized Milne-Pinney\cite{7} equation

\begin{equation}\label{5}\ddot{\rho}+\gamma(t)\dot{\rho}+\omega^2(t)\rho=\frac{1}{m^2(t)\rho^3}.\end{equation}

The invariant $I(t)$ satisfies the equation

\begin{equation}\label{6}\frac{dI}{dt}=\frac{\partial I}{\partial t}+\frac{1}{i\hbar}[I, H]=0\end{equation}and can be considered hermitian if we choose only the real solutions of Eq. (\ref{5}). Its eigenfunctions, $\phi_n(q,t)$, are assumed to form a complete orthonormal set with time-independent discrete eigenvalues, $\lambda_n$. Thus

\begin{equation}\label{7}I\phi_n(q, t)=\lambda_n\phi_n(q, t),\end{equation}with $\langle\phi_n,\phi_{n^\prime}\rangle=\delta_{nn^\prime}$.

Consider the Schr\"odinger equation (SE)

\begin{equation}\label{8}i\hbar\frac{\partial\psi(q, t)}{\partial t}=H(t)\psi(q, t),\end{equation}where $H(t)$ is given by Eq. (\ref{1}) with $p=-i\hbar\frac{\partial}{\partial q}$. Lewis and Riesenfeld\cite{6} showed that the solutions $\psi_n (q,t)$ of the SE (see Eq. (\ref{8})) are related to the functions $\phi_n (q,t)$ by

\begin{equation}\label{9}\psi_n(q, t)=e^{i\theta_n(t)}\phi_n(q, t),\end{equation}where the phase functions $\theta_n(t)$ satisfy the equation

\begin{equation}\label{10}\hbar\frac{d\theta_n(t)}{dt}=\langle\phi_n(q, t)|\left[i\hbar\frac{\partial}{\partial t}-H(t)\right]|\phi_n(q, t)\rangle.\end{equation}

The general solution of the SE (Eq. (\ref{8})) may be written as

\begin{equation}\label{11}\psi_n(q, t)=\sum_nc_ne^{i\theta_n(t)}\phi_n(q, t),\end{equation}where $c_n$ are time-independent coefficients.

Next, consider the unitary transformation

\begin{equation}\label{12}\phi_n^\prime(q, t)=\mathcal{U}\phi_n(q, t)\end{equation}where

\begin{equation}\label{13}\mathcal{U}=\exp{\left\{-i\left[\frac{m(t)\dot{\rho}}{2\hbar\rho}\right]q^2\right\}}.\end{equation}Under this transformation and defining $\sigma=q/\rho$, Eq. (\ref{7}) now reads

\begin{align}\label{14}I^\prime\varphi_n(\sigma)&=\left[-\left(\frac{\hbar^2}{2}\right)\frac{\partial^2}{\partial\sigma^2}+\left(\frac{\sigma^2}{2}\right)\right]\varphi_n(\sigma)\nonumber\\
&=\lambda_n\varphi_n(\sigma),\quad\lambda_n=\left(n+\frac{1}{2}\right)\hbar,\end{align}where $I^\prime=\mathcal{U}I\mathcal{U}^\dagger$ and $\frac{\varphi_n(\sigma)}{\rho^{1/2}}=\phi_n^\prime$. The factor $\rho^{1/2}$ warrants the normalization condition

\begin{equation}\label{15}\int{\phi_n^{\prime *}(q, t)\phi_n^\prime(q, t)}dq=\int{\varphi_n^{*}(q, t)\varphi_n(q, t)}d\sigma=1.\end{equation}

The solution of Eq. (\ref{14}) corresponds to that of the time-independent harmonic oscillator with $\lambda_n=(n+\frac{1}{2})\hbar$ . Then, by using Eqs. (\ref{12}), (\ref{13}) and (\ref{15}) we obtain

\begin{equation}\label{16}\phi_n(q,t)=\left[\frac{1}{\pi^{1/2}\hbar^{1/2}n!2^n\rho}\right]^{1/2}\exp{\left[\frac{im(t)}{2\hbar}\left(\frac{\dot{\rho}}{\rho}+\frac{i}{m(t)\rho^2}\right)q^2\right]}\times H_n\left[\left(\frac{1}{\hbar}\right)^{1/2}\frac{q}{\rho}\right],\end{equation}here $H_n$ is the usual Hermite polynomial of order $n$.

Applying $\mathcal{U}$ to the right-hand side of Eq. (\ref{10}) and after some algebra, we obtain

\begin{equation}\label{17}\theta_n(t)=-\left(n+\frac{1}{2}\right)\int_{t_0}^t{\frac{1}{m(t^\prime)\rho^2(t^\prime)}}dt^\prime.\end{equation}

Finally, using Eqs. (\ref{9}) and (\ref{16}) the exact solution of the SE for the time-dependent harmonic oscillator reads

\begin{equation}\label{18}\psi_n(q,t)=e^{i\theta_n(t)}\left[\frac{1}{\pi^{1/2}\hbar^{1/2}n!2^n\rho}\right]^{1/2}\exp{\left[\frac{im(t)}{2\hbar}\left(\frac{\dot{\rho}}{\rho}+\frac{i}{m(t)\rho^2}\right)q^2\right]}\times H_n\left[\left(\frac{1}{\hbar}\right)^{1/2}\frac{q}{\rho}\right].\end{equation}

\section{\label{sec3}WAVE FUNCTIONS OF TIME-DEPENDENT LOG-PERIODIC OSCILLATORS}

In Ref. \onlinecite{5}, \"{O}zeren considered five different variations of $m(t)$ and $k(t)$, namely: (i) $m(t)=m_0$ and $k(t)=k_0\left(\frac{t_0}{t}\right)^2$; (ii) $m(t)=m_0 \left(\frac{t}{t_0}\right)^2$ and $k(t)=k_0$; (iii) $m(t)=m_0\left(\frac{t}{t_0}\right)^\alpha$ and $k(t)=k_0 \left(\frac{t_0}{t}\right)^{(\alpha+2)}$; (iv) $m(t)=m_0 \left(\frac{t}{t_0}\right)$ and $k(t)=k_0 \left(\frac{t_0}{t}\right)$; and (v) $m(t)=m_0 \left(\frac{t}{t_0}\right)^\alpha$ and $k(t)=k_0\left(\frac{t}{t_0}\right)^\alpha$. Here we consider only three ((i), (ii) and (iv)) out of the five oscillators studied by \"{O}zeren\cite{5}, for which $\omega(t)=\sqrt{\frac{k(t)}{m(t)}}=\omega_0\frac{t_0}{t}$.

\subsection{$\bm{m(t)=m_0\frac{t}{t_0}}$ and $\bm{k(t)=k_0\frac{t_0}{t}}$}

In this case Eqs. (\ref{2}) and (\ref{5}) read

\begin{equation}\label{19}\ddot{q}+\frac{1}{t}\dot{q}+\frac{\omega_0^2t_0^2}{t^2}q=0\end{equation}and

\begin{equation}\label{20}\ddot{\rho}+\frac{1}{t}\dot{\rho}+\frac{\omega_0^2t_0^2}{t^2}\rho=\frac{t_0^2}{m_0^2}\frac{1}{t^2\rho^3},\end{equation}respectively.

Following the procedure described in Ref.\onlinecite{7}, we find  $\rho=c=\frac{1}{\sqrt{m_0\omega_0}}$ . From Eqs. (\ref{17}) and (\ref{18}) we have

\begin{equation}\label{21}\psi_n(q,t)=e^{-i\left(n+\frac{1}{2}\right)\omega_0t_0\ln{\frac{t}{t_0}}}\left[\frac{m_0\omega_0}{\pi\hbar(n!)^22^{2n} }\right]^{1/4}\exp{\left[-\frac{m_0^2\omega_0^2q^2}{2\hbar}\right]}\times H_n \left[\left(\frac{m_0\omega_0}{\hbar}\right)^{1/2}q \right],\end{equation}which, except for the phase factor, is similar to the well-known wave function for the time-independent harmonic oscillator. 

The coherent states for the time-dependent harmonic oscillator (Eq.(\ref{1})) are constructed as follows\cite{9}. Consider the time-dependent creation ($a^\dagger(t)$) and annihilation ($a(t)$) operators defined as

\begin{equation}\label{22}a^\dagger(t)=\left(\frac{1}{2\hbar}\right)^{1/2}\left[ \left(\frac{q}{\rho}\right)-i(\rho p-m\dot{\rho}q)\right]\end{equation}

\begin{equation}\label{23}a(t)=\left(\frac{1}{2\hbar}\right)^{1/2}\left[ \left(\frac{q}{\rho}\right)+i(\rho p-m\dot{\rho}q)\right],\end{equation}where $[a^\dagger(t),a(t)]=1$. In terms of $a(t)$ and $a^\dagger(t)$ the invariant $I$ (see Eq. (\ref{4})) can be written as

\begin{equation}\label{24}I=\hbar\left(a^\dagger(t)a(t)+\frac{1}{2}\right).\end{equation}

Let $|n,t\rangle$ be the eigenstates of $I$. Therefore the following relations hold

\begin{equation}\label{25}a(t)=\sqrt{n}|n-1,t\rangle\end{equation}

\begin{equation}\label{26}a^\dagger(t)=\sqrt{n+1}|n+1,t\rangle,\end{equation}

\begin{equation}\label{27}I|n,t\rangle=\hbar\left(n+\frac{1}{2}\right)|n,t\rangle.\end{equation}

Since the coherent states for $I$ can be easily constructed, the coherent states for the time-dependent harmonic oscillator are straightforwardly obtained:

\begin{equation}\label{28}|\alpha,t\rangle=e^{-|\alpha|^2/2} \sum_{n=0}^\infty{\frac{\alpha^n}{(n!)^{1/2}}e^{i\theta_n(t)} |n,t\rangle},\end{equation}where $\theta_n$ is given by Eq. (\ref{17}), and the complex number $\alpha(t)$ satisfies the eigenvalue equation

\begin{equation}\label{29}a(t)|\alpha,t\rangle=\alpha(t)|\alpha,t\rangle,\end{equation}with

\begin{equation}\label{30}\alpha(t)=\alpha(t_0)e^{2i\theta_0(t)}\end{equation}and

\begin{equation}\label{31}\theta_0(t)=-\frac{1}{2}\int_{t_0}^t\frac{dt^\prime}{m(t^\prime)\rho^2(t^\prime)}.\end{equation}

The fluctuations in $q$ ($\Delta q$) and $p$ ($\Delta p$) and the uncertainty product ($\Delta q\Delta p$) in the coherent state $|\alpha,t\rangle$ , read

\begin{equation}\label{32}\Delta q_\alpha=\sqrt{\langle q^2\rangle_\alpha-\langle q\rangle_\alpha^2}=\sqrt{\frac{\hbar}{2}}\rho,\end{equation}

\begin{equation}\label{33}\Delta p_\alpha=\sqrt{\langle p^2\rangle_\alpha-\langle p\rangle_\alpha^2}=\sqrt{\frac{\hbar}{2}}\frac{1}{\rho}\left(1+m^2\dot{\rho}^2\rho^2\right)^{1/2}\end{equation}and

\begin{equation}\label{34}\Delta q_\alpha\Delta p_\alpha=\frac{\hbar}{2}\left(1+m^2\dot{\rho}^2\rho^2\right)^{1/2},\end{equation}respectively.

If $m(t)\dot{\rho}\rho\neq0$, $\Delta q_\alpha\Delta p_\alpha$ is not minimum, indicating that the coherent states $|\alpha,t\rangle$ are not minimum-uncertainty (coherent) states. In fact, the states $|\alpha,t\rangle$ for the time-dependent harmonic oscillator are equivalent to the well-known squeezed states, as pointed out in Refs. \onlinecite{10, 11}.

For $\rho=c$, $\dot{\rho}=0$ and $\Delta q_\alpha\Delta p_\alpha=\frac{\hbar}{2}$ , indicating that the states $|\alpha,t\rangle$ are ``true" coherent states.  This is an interesting result since the minimum uncertainty product is assumed to be satisfied only for time-independent harmonic oscillator, unless the solution of Eq.(\ref{5}) is a constant\cite{1}.
 
Next, let us analyze the time behavior of $\langle q\rangle_\alpha$, $\langle p\rangle_\alpha$ and the phase diagram $\langle q\rangle_\alpha\times\langle p\rangle_\alpha$. By setting $\alpha(t_0 )=u+iv$ and using Eqs. (\ref{22}) and (\ref{23}), we find

\begin{equation}\label{35}\langle q\rangle_\alpha=\sqrt{2\hbar}\left[u\cos{\left(\theta_0(t)\right)}-v\sin{\left(\theta_0(t)\right)}\right]\end{equation}

\begin{equation}\label{36}\langle p\rangle_{\alpha}=\sqrt{2\hbar}\left[\left(\frac{v}{\rho}+um\dot{\rho}\right)\cos{\left(\theta_0(t)\right)}+\left(\frac{u}{\rho}-vm\dot{\rho}\right)\sin{\left(\theta_0(t)\right)}\right].\end{equation}

The constants $u$ and $v$ are determined from the initial conditions $\langle q(t_0)\rangle_\alpha=q_0$ and $\langle p(t_0)\rangle_\alpha=p_0=m(t_0)v_0$. For $q_0=1$ and $v_0=0$, we find

\begin{equation}\label{37}\langle q\rangle_\alpha=\cos{\left(t_0\omega_0\ln{\frac{t}{t_0}}\right)},\end{equation}

\begin{equation}\label{38}\langle p\rangle_\alpha=-m_0\omega_0\sin{\left(t_0\omega_0\ln{\frac{t}{t_0}}\right)}.\end{equation}

\begin{figure}[t]
	\centering
		\includegraphics{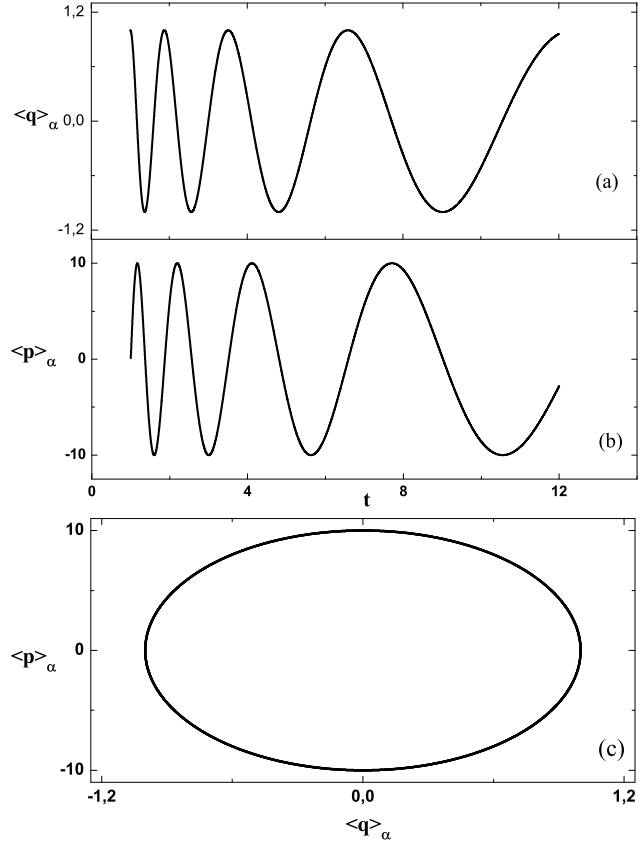}
	\caption{Plots of  (a) $\langle q\rangle_\alpha$, (b) $\langle p\rangle_\alpha$, and (c) the phase diagram $\langle p\rangle_\alpha$ vs $\langle q\rangle_\alpha$. In the plots we used $t_0=1.0$, $q_0=1.0$, $v_0=0.0$, $\omega_0=10.0$ and $m_0=1.0$.}
	\label{fig1}
\end{figure}

Figures \ref{fig1}(a) and (b) show the time dependent behavior of $\langle q\rangle_\alpha$ and $\langle p\rangle_\alpha$, respectively. In all plots we used $t_0=1.0$, $\omega_0=10.0$ and $m_0=1.0$.  From Fig. \ref{1}(a) we observe that the system oscillates forth and back between the classical turning points with an increasing period and constant amplitude. The phase diagram is shown in Fig. \ref{1}(c). Even though that this system is dissipative (total energy $E=\frac{1}{2t}$), it behaves like the usual time-independent harmonic oscillator ($E=$ constant). This can be seen from the relation $A=\sqrt{\frac{2E}{k}}$, where $A$ is the amplitude of motion. Since $k\propto\frac{1}{t}$  and $E\propto\frac{1}{t}$, $A$ is a constant. As $t$ increases, the frequency $\omega(t)$ decreases ($\propto\frac{1}{t}$) while the period increases ($\propto\frac{t}{lnt}$) leading to the ``exact" log periodic behavior shown in Fig. \ref{1}(a).

Pedrosa et al\cite{12} have combined linear invariants and the LR method to obtain the exact wave function for a particle trapped by oscillating fields, which were written in terms of Mathieu functions.  They calculated  $\Delta q\Delta p$ and the quantum correlation between $q$ and $p$, defined by $C_{1,1}=\frac{1}{2}\langle\left(qp+pq\right)\rangle-\langle q\rangle\langle p\rangle$\cite{13}. They are related through the equation

\begin{equation}\label{39}\Delta q\Delta p=\frac{\hbar}{2}\sqrt{1+\left(\frac{2}{\hbar}C_{1,1}\right)^2},\end{equation}which shows that  $\Delta q\Delta p$ is minimum whenever $C_{1,1}=0$, as it happens for  $C_{1,1}$ calculated in the coherent state $|\alpha,t\rangle$, i.e., $(C_{1,1})_\alpha=0$. The fact that $C_{1,1}=0$ does not mean that $q$ and $p$ are uncorrelated. In order to verify the correlation between  $q$ and $p$ one may study the function $C_{n,m}=\frac{1}{2}\langle\left(q^np^m+p^mq^n\right)\rangle-\langle q^n\rangle\langle p^m\rangle$. For the coherent state $|\alpha,t\rangle$, we find that  $(C_{2,2} )_\alpha=-\frac{\hbar^2}{2}$ , indicating that $q$ and $p$ even assumed as ``classical" quantities are correlated. 

The uncertainty product and correlations in the state $\psi_n$ (Eq.(\ref{21})) are more easily calculated using the relation $|\psi_n(q,t)\rangle=e^{i\theta_n(t)}|n,t\rangle$. They are given by

\begin{equation}\label{40}\Delta q_{\psi_n}\Delta p_{\psi_n}=\left(n+\frac{1}{2}\right)\hbar,\end{equation}

\begin{equation}\label{41}\left(C_{1,1}\right)_{\psi_n}=0\end{equation}and

\begin{equation}\label{42}\left(C_{2,2}\right)=-\left(n^2+n+\frac{1}{2}\right)\hbar^2.\end{equation}

We noticed that Eq. (\ref{39}) is satisfied for $n=0$. In this case  $\psi_0$ given by

\begin{equation}\label{43}\psi_0(q,t)=e^{-i\frac{t_0\omega_0}{2}\ln{\left(\frac{t}{t_0}\right)}}\left(\frac{m_0\omega_0}{\pi\hbar}\right)^{1/4}\exp{\left[-\frac{m_0^2\omega_0^2 q^2}{2\hbar}\right]}\end{equation}is the coordinate representation of the coherent state\cite{14}.

\subsection{$\bm{m(t)=m_0}$ and $\bm{k(t)=k_0\left(\frac{t_0}{t}\right)^2}$}

In this case Eqs. (\ref{2}) and (\ref{5}) are, respectively, given by

\begin{equation}\label{44}\ddot{q}+\omega_0^2\frac{t_0^2}{t^2}q=0\end{equation}and

\begin{equation}\label{45}\ddot{\rho}+\omega_0^2\frac{t_0^2}{t^2}\rho=\frac{1}{m_0\rho^3}.\end{equation}

Following the procedure described in Refs. \onlinecite{7, 8}, we find $\rho=\sqrt{\frac{2}{m_0}}\frac{\sqrt{t}}{\left(4\omega_0^2 t_0^2-1\right)^{1/4}}$ and from Eqs. (\ref{17}) and (\ref{18}) we have

\begin{align}\label{46}\psi_n(q,t)=&e^{-\frac{i}{2}\left(n+\frac{1}{2}\right)\left(4\omega_0^2t_0^2-1\right)^{1/2}\ln\left(\frac{t}{t_0}\right)}\left[\frac{m_0\left(4\omega_0^2t_0^2-1\right)^{1/2}}{\pi\hbar(n!)^22^{2n+1}}\right]^{1/4}\times\frac{1}{t^{1/4}}\nonumber\\ \times&\exp{\left\{\frac{m_0}{4\hbar t}\left[i-\left(4\omega_0^2t_0^2-1\right)^{1/2}\right]q^2\right\}}
\times H_n\left[\left(\frac{m_0}{2\hbar}\right)^{1/2}\frac{\left(4\omega_0^2t_0^2-1\right)^{1/4}}{\sqrt{t}}q\right]\end{align}

The values of  $\Delta q\Delta p$ and  $C_{1,1}$ in the state  $|\psi_n (q,t)\rangle$, are given by, respectively

\begin{equation}\label{47}\Delta q_{\psi_n}\Delta p_{\psi_n}=\frac{2\omega_0 t_0}{\left(4\omega_0^2t_0^2-1\right)^{1/2}}\left(n+\frac{1}{2}\right)\hbar,\end{equation}and

\begin{equation}\label{48}\left(C_{1,1}\right)_{\psi_n}=-\left(\frac{1}{4\omega_0^2t_0^2-1}\right)^{1/2}\left(n+\frac{1}{2}\right)\hbar.\end{equation}

For $n=0$, $\Delta q_{\psi_0}\Delta p_{\psi_0}=\frac{\omega_0 t_0}{\left(4\omega_0^2t_0^2-1\right)^{1/2}}\hbar$, and the state

\begin{align}\label{49}\psi_0(q,t)=e^{-\frac{i}{4}\left(4\omega_0^2t_0^2-1\right)^{1/2}\ln{\left(\frac{t}{t_0}\right)}}&\left[\frac{m_0\left(4\omega_0^2t_0^2-1\right)^{1/2}}{2\pi\hbar}\right]^{1/4}\times\frac{1}{t^{1/4}}\nonumber\\
\times&\exp{\left\{\frac{m_0}{4\hbar t}\left[i-\left(4\omega_0^2t_0^2-1\right)^{1/2}q^2\right]\right\}}\end{align}corresponds to the coordinate representation of the squeezed state\cite{14}.

For the sake of comparison with case \textbf{A}, let us discuss the behavior of the classical variables $q$ and $p$ on time, as well as the phase diagram. By solving Eq.(\ref{44}), the  solutions for $q$ and $p$ satisfying the initial conditions $q_0=1$ and $v_0=0$ are, respectively, given by

\begin{equation}\label{50}q(t)=\sqrt{\frac{t}{t_0}}\left[\cos{\left(\frac{\left(4\omega_0^2t_0^2-1\right)^{1/2}}{2}\ln{\frac{t}{t_0}}\right)}-\frac{1}{\left(4\omega_0^2t_0^2-1\right)^{1/2}}\sin{\left(\frac{\left(4\omega_0^2t_0^2-1\right)^{1/2}}{2}\ln{\frac{t}{t_0}}\right)}\right]\end{equation}

And

\begin{equation}\label{51}p(t)=-\sqrt{\frac{t_0}{t}}\frac{2m_0\omega_0^2t_0}{\left(4\omega_0^2t_0^2-1\right)^{1/2}}\sin{\left(\frac{\left(4\omega_0^2t_0^2-1\right)^{1/2}}{2}\ln{\frac{t}{t_0}}\right)}.\end{equation}

\begin{figure}[t]
	\centering
		\includegraphics{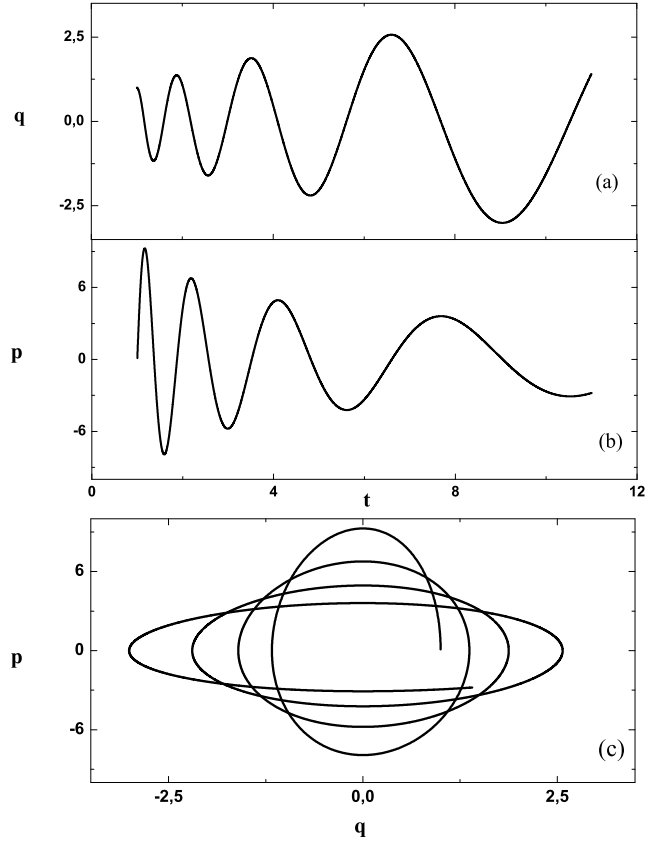}
	\caption{Plots of  (a) q, (b) p, and (c) the phase diagram  $p$ vs $q$. In the plots we used $t_0=1.0$, $q_0=1.0$, $v_0=0.0$, $\omega_0=10.0$ and $m_0=1.0$.}
	\label{fig2}
\end{figure}

Figures \ref{fig2}(a) and \ref{fig2}(b) show the variation of $q$ and $p$ on time, respectively. Unlike from case \textbf{A} where the system oscillates back and forth between the turning points with constant amplitude, here $q$ increases while that $p$ decreases as time increases. Figure \ref{fig2}(c) shows the phase diagram for this oscillator. Initially at rest, the particle is speeded up, and then slowed down, indicating that the system is also dissipative. Since $E\propto 1/t$ and $k\propto 1/t^2$, the amplitude $A$ increases as $A\propto \sqrt{t}$. Due to the presence of the factor $\sqrt{t}$ in Eq. (\ref{50}) this oscillator exhibits a pseudo-log-periodic behavior.

\subsection{$\bm{m(t)=m_0\left(\frac{t}{t_0}\right)^2}$ and $\bm{k(t)=k_0$}}

Now Eqs. (\ref{2}) and (\ref{5}) are, respectively, given by

\begin{equation}\label{52}\ddot{q}+\frac{2}{t}\dot{q}+\omega_0^2\frac{t_0^2}{t^2}q=0\end{equation}and

\begin{equation}\label{53}\ddot{\rho}+\frac{2}{t}\dot{\rho}+\omega_0^2\frac{t_0^2}{t^2}\rho=\frac{t_0^4}{m_0^2}\frac{1}{t^4\rho^3}.\end{equation}

Again, by following the procedure described in Refs. \onlinecite{7, 8}, we find $\rho=\sqrt{\frac{2}{m_0}}\frac{t_0}{\left(4\omega_0^2 t_0^2-1\right)^{1/4}}\frac{1}{\sqrt{t}}$ and from Eqs. (\ref{17}) and (\ref{18}) we obtain

\begin{align}\label{54}\psi_n(q,t)=&e^{-\frac{i}{2}\left(n+\frac{1}{2}\right)\frac{\left(4\omega_0^2t_0^2-1\right)^{1/2}}{t_0}\ln\left(\frac{t}{t_0}\right)}\left[\frac{m_0\left(4\omega_0^2t_0^2-1\right)^{1/2}}{\pi\hbar t_0(n!)^22^{2n+1}}\right]^{1/4}\times t^{1/4}\nonumber\\ \times&\exp{\left\{\frac{m_0t}{4\hbar }\left[i-\frac{\left(4\omega_0^2t_0^2-1\right)^{1/2}}{t_0}\right]q^2\right\}}
\times H_n\left[\left(\frac{m_0}{2\hbar t_0}\right)^{1/2}\left(4\omega_0^2t_0^2-1\right)^{1/4}\sqrt{t}q\right]\end{align}

The values of  $\Delta q\Delta p$ and  $C_{1,1}$ in the state  $|\psi_n (q,t)\rangle$, are respectively given by

\begin{equation}\label{55}\Delta q_{\psi_n}\Delta p_{\psi_n}=\frac{2\omega_0 t_0}{\left(4\omega_0^2t_0^2-1\right)^{1/2}}\left(n+\frac{1}{2}\right)\hbar\end{equation}and

\begin{equation}\label{56}\left(C_{1,1}\right)_{\psi_n}=-\left(\frac{1}{4\omega_0^2t_0^2-1}\right)^{1/2}\left(n+\frac{1}{2}\right)\hbar.\end{equation}

The expression of the coordinate representation of the squeezed state for $n=0$, reads

\begin{align}\label{57}\psi_0(q,t)=e^{-\frac{i}{4}\frac{\left(4\omega_0^2t_0^2-1\right)^{1/2}}{t_0}\ln{\left(\frac{t}{t_0}\right)}}&\left[\frac{m_0\left(4\omega_0^2t_0^2-1\right)^{1/2}}{2\pi\hbar t_0^2}\right]^{1/4}\times\frac{1}{t^{1/4}}\nonumber\\
\times&\exp{\left\{\frac{m_0t}{4\hbar }\left[i-\frac{\left(4\omega_0^2t_0^2-1\right)^{1/2}}{t_0}\right]q^2\right\}}.\end{align}

By solving Eq. (\ref{52}) and using the initial conditions $q_0=1$ and $v_0=0$, we find

\begin{equation}\label{58}q(t)=\sqrt{\frac{t_0}{t}}\left[\cos{\left(\frac{\left(4\omega_0^2t_0^2-1\right)^{1/2}}{2}\ln{\frac{t}{t_0}}\right)}-\frac{1}{\left(4\omega_0^2t_0^2-1\right)^{1/2}}\sin{\left(\frac{\left(4\omega_0^2t_0^2-1\right)^{1/2}}{2}\ln{\frac{t}{t_0}}\right)}\right]\end{equation}and

\begin{equation}\label{59}p(t)=\sqrt{\frac{t}{t_0}}\frac{2m_0\omega_0^2t_0}{\left(4\omega_0^2t_0^2-1\right)^{1/2}}\sin{\left(\frac{\left(4\omega_0^2t_0^2-1\right)^{1/2}}{2}\ln{\frac{t}{t_0}}\right)}.\end{equation}

\begin{figure}[t]
	\centering
		\includegraphics{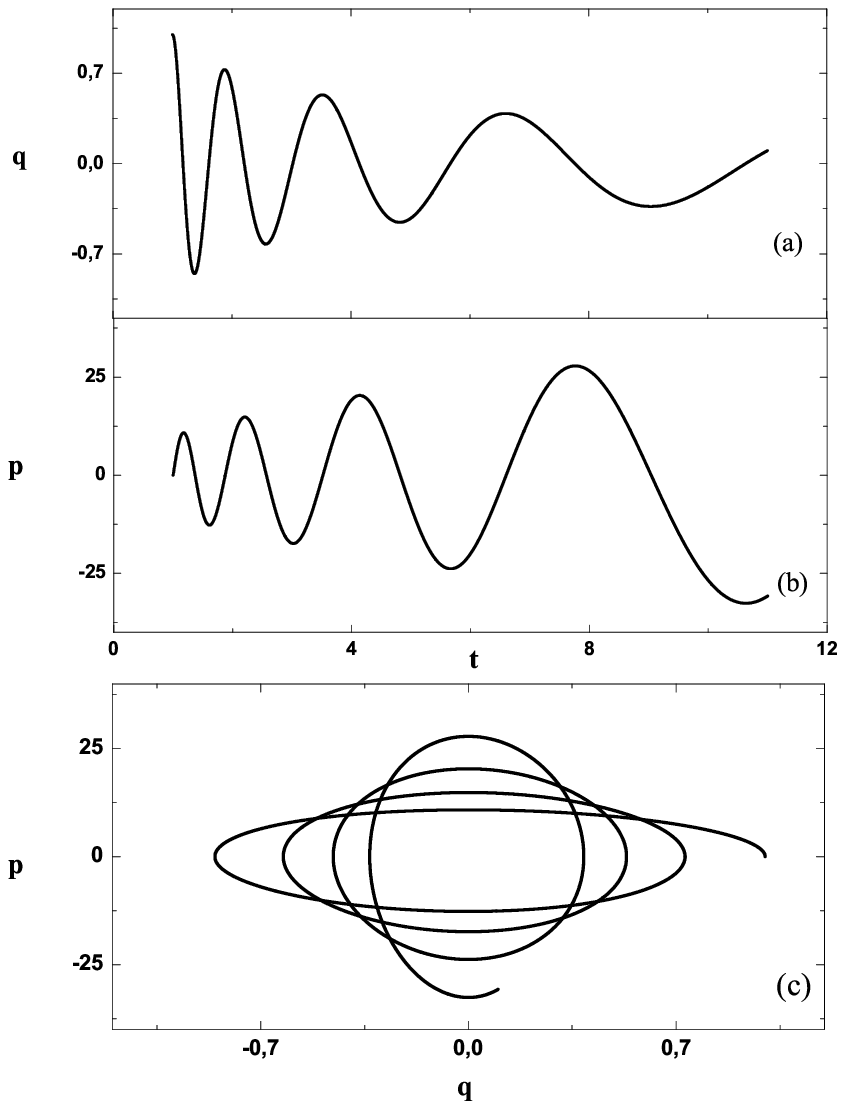}
	\caption{Plots of  (a) $q$, (b) $p$, and (c) the phase diagram $p$ vs $q$. In the plots we used $t_0=1.0$, $q_0=1.0$, $v_0=0.0$, $\omega_0=10.0$ and $m_0=1.0$.}
	\label{fig3}
\end{figure}

The behavior of the classical $q$ and $p$ variables on time is displayed in Figs. \ref{fig3} (a) and (b), respectively. Despite the oscillating ($\cos{\left(\frac{\sqrt{3}}{2}\ln{t}\right)}$ and ($\sin{\left(\frac{\sqrt{3}}{2}\ln{t}\right)}$) terms, $q$ and $p$ exhibit an opposite behavior compared to those calculated in case \textbf{B}. Here $q$ decreases while $p$  increases as $t$ increases. Figure \ref{fig3}(c) shows the phase diagram. We observe that the amplitude $A$ decreases as $\propto 1/\sqrt{t}$. This oscillator also exhibits a pseudo-log-periodic character.

\section{\label{sec4}CONCLUDING REMARKS}

In this paper we have used a unitary transformation and the LR invariant method in the Schr\"{o}dinger picture to obtain the exact wave functions for oscillators exhibiting either log-periodic or pseudo-log-periodic behavior. It is well-known that a challenge in obtaining the exact solution (see Eq. (\ref{18})) for the SE (Eq.(\ref{8})) for $H(t)$ given in Eq. (\ref{1}), is the solution of the auxiliary equation for the c-number quantity $\rho$ (see Eq. (\ref{5})). Here we find $\rho$ for each case using the methods described in Refs. \onlinecite{7, 8}.  

For case $A$, we find $\rho=c$ and, as a consequence, the solution for $\psi_n (q,t)$ (see Eq. (\ref{21})) except for the phase factor ($e^{-i\left(n+\frac{1}{2}\right)t_0\omega_0\ln{\frac{t}{t_0}}}$), is similar to the well-known wave function for the time-independent harmonic oscillator of mass $m_0$ and frequency $\omega_0$. In Ref. \onlinecite{1}, we observed that when $m(t)=m_0$, $\omega(t)=\omega_0$, and  $\rho(t)=\left(\frac{1}{m_0\omega_0}\right)^{1/2}$, which is a particular solution of Eq. (\ref{5}), the wave function obtained also corresponds to that of the time-independent harmonic oscillator. In case \textbf{A} even with $m\propto t$ and $\omega\propto \frac{1}{t}$, we obtain the same solution for $\rho$ ($\rho= c$), indicating that this oscillator behaves as the harmonic oscillator with $m$ and $\omega$ constant.

We have constructed the ``true" coherent states, $|\alpha,t\rangle$, whose coordinate representation is given by Eq. (\ref{43}). We verified that Eq. (\ref{39}) holds for $|\alpha,t\rangle$. We calculated the quantum fluctuations in the coordinate and momentum as well as the quantum correlations between the coordinate and momentum in the state $\psi_n (q,t)$. 

We analyzed the time behavior of $\langle q\rangle_\alpha$ and $\langle p\rangle_\alpha$ , as well as the phase diagram $\langle q\rangle_\alpha\times\langle p\rangle_\alpha$ (see Fig.\ref{fig1}(a)-(c)). We observed that $\langle q\rangle_\alpha$ and $\langle p\rangle_\alpha$ exhibits the exact log-periodic behavior, and that the phase diagram indicates, as already anticipated, that the log-periodic oscillator behaves as the classical harmonic oscillator with $m(t)=m_0$ and $\omega(t)=\omega_0$.  

For cases \textbf{B} and \textbf{C}, we obtained the wave functions given by Eqs. (\ref{46}) and (\ref{54}), respectively.

\end{document}